\documentclass[square,aip,preprint,showkeys,superscriptaddress]{revtex4}
\usepackage{graphicx}% Include figure files
\usepackage{epstopdf,amsmath}

\begin{document}
\title{Proton irradiation effects on metal-YBCO interfaces}
\author{C. Acha}
\thanks{corresponding author (acha@df.uba.ar)}
\affiliation{Universidad de Buenos Aires, Facultad de Ciencias Exactas y Naturales, Departamento de  F\'{\i}sica, Laboratorio de Bajas Temperaturas and IFIBA,
    UBA-CONICET, Pabell\'on I, Ciudad Universitaria, C1428EHA CABA, Argentina}
\author{G. A. Sanca}
\affiliation{Escuela de Ciencia y Tecnolog\'{\i}a, Universidad Nacional de San Mart\'{\i}n,
	Mart\'{\i}n de Irigoyen 3100, B1650JKA San Mart\'{\i}n, Bs. As.,
	Argentina}
\author{M. Barella}
\thanks{Present address: CIBION, C1428EHA CABA, Argentina}
\affiliation{Escuela de Ciencia y Tecnolog\'{\i}a, Universidad Nacional de San Mart\'{\i}n,
	Mart\'{\i}n de Irigoyen 3100, B1650JKA San Mart\'{\i}n, Bs. As.,
	Argentina} \affiliation{Consejo Nacional de Investigaciones
Cient\'{\i}ficas y T\'ecnicas (CONICET), Godoy Cruz 2290, C1425FQB
CABA, Argentina}
\author{M. Alurralde}
\affiliation{Comisi\'on Nacional de Energ\'{\i}a At\'omica (CNEA),
Av. Del Libertador 8250, C1429BNP, Argentina}
\author{F. Gomez Marlasca}
\affiliation{Comisi\'on Nacional de Energ\'{\i}a At\'omica (CNEA),
	Av. Del Libertador 8250, C1429BNP, Argentina}
\author{H. Huhtinen} 
\affiliation{University of Turku, Department of Physics and Astronomy, Wihuri Physical Laboratory, FI-20014 Turku, Finland}
\author{P. Paturi}
\affiliation{University of Turku, Department of Physics and Astronomy, Wihuri Physical Laboratory, FI-20014 Turku, Finland}
\author{F. Golmar} 
\affiliation{Escuela de Ciencia y Tecnolog\'{\i}a, Universidad Nacional de San Mart\'{\i}n,
	Mart\'{\i}n de Irigoyen 3100, B1650JKA San Mart\'{\i}n, Bs. As.,
	Argentina} \affiliation{Consejo Nacional de Investigaciones
Cient\'{\i}ficas y T\'ecnicas (CONICET), Godoy Cruz 2290, C1425FQB
CABA, Argentina}
\author{P. Levy}
	\affiliation{Escuela de Ciencia y Tecnolog\'{\i}a, Universidad Nacional de San Mart\'{\i}n,
	Mart\'{\i}n de Irigoyen 3100, B1650JKA San Mart\'{\i}n, Bs. As.,
	Argentina}
	\affiliation{Consejo
	Nacional de Investigaciones Cient\'{\i}ficas y T\'ecnicas (CONICET),
	Godoy Cruz 2290, C1425FQB CABA, Argentina} 
	\affiliation{Comisi\'on Nacional de Energ\'{\i}a At\'omica (CNEA),
		Av. Del Libertador 8250, C1429BNP, Argentina} 
\date{\today}
\draft

\begin{abstract}
10 MeV proton-irradiation effects on a YBCO-based test structure were analyzed by measuring its current-voltage (IV) characteristics for different cumulated fluences. For fluences of up to $\sim$~80$\cdot$10$^9$~p/cm$^2$ no changes in the electrical behavior of the device were observed, while for a fluence of $\sim$~300$\cdot$10$^9~$ p/cm$^2$ it becomes less conducting. A detailed analysis of the room temperature IV characteristics based on the $\gamma$ power exponent parameter [$\gamma=dLn(I)/dLn(V)$] allowed us to reveal the main conduction mechanisms as well as to establish the equivalent circuit model of the device. The changes produced in the electrical behavior, in accordance with Monte Carlo TRIM simulations, suggest that the main effect induced by protons is the displacement of oxygen atoms within the YBCO lattice, particularly from oxygen-rich to oxygen-poor areas, where they become trapped. 
\end{abstract}


\maketitle

\section{INTRODUCTION}

There are many applications where it is of vital importance to have reliable non-volatile memories (NVM) that withstand the harsh conditions the environment impose. This is the case in spaceborne applications, where the use of radiation tolerant, thermal cycling resistant memories is essential to keep critical information uncorrupted, both for navigation and operation purposes~\cite{Gerardin10}. Another relevant example are nuclear power plants where robotic surveillance and maintenance in high radiation zones is crucial for safety~\cite{Pegman06,Yoshida14}.

As Flash memories are very sensitive to both Single Event Effects and cumulative effects like Total Ionizing Dose (TID) or Displacement Damage (DD)~\cite{Gerardin13}, the emerging Resistive Random Access Memories (ReRAM) technology is a promising candidate to be used as NVM under such severe conditions. This new type of memories have shown to be tolerant against different types of ionizing radiation~\cite{Chen14,Velo17,Bi18}.

The general conclusions on their functionality under irradiation indicate that metal-oxide-metal structures withstand large doses of ionizing radiation while maintaining operation, within a reliability margin. For example, Barnaby {\textit{et al.}}~\cite{Barnaby13} exposed TiO$_{\mathrm{2}}$ based-devices to fluences of up to 10$^{14}$/cm$^{2}$ of 1~MeV alpha particles without sensing significant changes on their configured states. Hughart {\textit{et al.}}~\cite{Hughart13} studied radiation effects on TaO$_{\mathrm{x}}$ and TiO$_{\mathrm{2}}$-based devices and compared their performance. They reported that both technologies exhibit changes in their resistance state when exposed to fluences of up to 10$^{10}$/cm$^{2}$ of 800 keV Ta ions. However, after irradiation, devices continued to operate normally and even recovered their resistance states with repeated switching cycles.

Among all possible ReRAM candidates, those based on the valence change (VC) produced by electric field-driven diffusion of oxygen vacancies~\cite{Waser09a,Rozenberg10} represent an interesting choice, for different reasons. On the one hand, as their interfacial resistivity depends mostly on the volume properties of the oxide rather that on surface properties of the interface, they may be less sensitive to ionizing radiation defects. On the other hand, if defects introduced by the ionizing radiation affect the relevant microscopic parameters associated with the specific electric transport mechanisms through an interface and its related oxide volume (like trap energy, density of traps, work function, carrier mobility, etc), the electrical changes can be used to understand the irradiation damage. In this way, the analysis of the macroscopic electrical response may provide an attainable and useful tool when introduced defects are hard to detect by usual electronic microscopies.

Although being mostly known for its superconducting properties at T $\leq$ T$_c$ ($\simeq$ 90 K),  YBa$_2$Cu$_3$O$_{7-\delta}$ (YBCO) has been widely used to study their resistive switching properties at room temperature  for ReRAM memory applications~\cite{Acha09a,Acha09b,Placenik10,Acha11,Lanosa20}. What makes YBCO such an interesting material to build ReRAM devices is the possibility of modifying its oxygen content by applying moderate electric fields and thus producing large changes in its conductivity~\cite{Moeckly93,Schulman13,Palau18}. As will be discussed in detail later, transport properties through metal-YBCO interfaces are controlled by oxygen content inhomogeneities of YBCO, close to the interface~\cite{Schulman12,Placenik12,Schulman15,Waskiewicz15,Truchly16,Waskiewicz18,Tulina18}.  Additionally, YBCO-based memories were successfully used onboard a small satellite orbiting at a Low Earth Orbit (LEO) over a period of more than one year~\cite{Acha20}, despite the hostile conditions imposed by lift-off vibrations, permanent thermal cycling ($\pm$ 10 $^{\circ}$C 16$\times$/day) and an ionizing radiation representing a cumulative dose 3 orders of magnitude greater than the one they would have received in the same period of time on the surface of the Earth. As one of the most important sources of radiation effects at LEO is related to protons, the main motivation of this paper is to determine the radiation tolerance of these memories (that is, the maximum proton fluence at which alterations in their electrical behavior are not yet apparent). To do so, the underlying transport mechanisms of metal-YBCO interfaces were studied over a cumulative series of 10 MeV proton irradiations.   

In this article, we analyze the effects of 10 MeV protons on the current-voltage (IV) characteristics of a Au-YBCO based ReRAM memory device. No changes were detected for fluences of up to 80$\cdot$10$^9$ p/cm$^2$, while for 295$\cdot$10$^9$ p/cm$^2$ the device under test (DUT) became less conducting. A detailed analysis of the electrical conduction mechanisms through this interface enabled us to model an equivalent schematic circuit and, more importantly, to infer the microscopic effects produced by the irradiation with protons. Monte Carlo Transport of Ion in Matter (TRIM) simulations confirm our interpretation of the data, which indicates that protons mainly knock-on oxygen atoms out of the YBCO structure transferring them from the oxygen-rich zones to oxygen-poor ones.

\section{EXPERIMENTAL DETAILS}

YBCO thin films were grown by using the pulsed laser deposition
(PLD) technique. Fully relaxed films were obtained on top of a 500 $\mu$m thick (100) single crystal SrTiO$_3$ substrate.  A 150 nm YBCO thick layer was deposited by applying 1500 pulses with a growth rate of 0.1 nm/pulse, as determined by previous calibrations~\cite{Khan19}. Resistivity and magnetization measurements confirmed that the YBCO films are near optimally-doped and have a superconducting transition temperature at $\sim$ 90 K. Additional details of their synthesis and characterization can be found elsewhere~\cite{Huhtinen01,Paturi04,Peurla07,Khan19}.

\subsection{Electrical characterizations}

\begin{figure}[htp]
	\vspace{-0mm}
	\centerline{\includegraphics[width=6.5in]{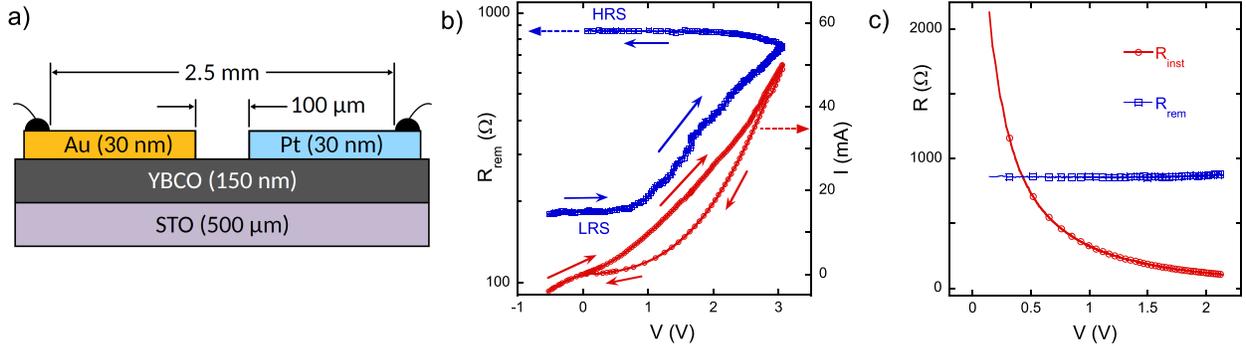}}
	%\centerline{\includegraphics[angle=0,scale=0.2]{R4wycontactos4.eps}}
	\vspace{-0mm}\caption{(Color online) a) Schematic represention of the cross section of the DUT (not to scale). b) Remanent resistance R$_{rem}$ (blue) representing half of a resistance hysteresis switching loop (RHSL) and IV characteristics (red) of the DUT during the pulsing treatment to induce a reset. c) After the reset process, R$_{inst}$ (red) shows the non-linear dependence upon increasing the amplitude of the voltage pulses (V) while R$_{rem}$ (blue) remains constant, indicating that no resistance switching (RS) occurs during a sweep of a lower voltage amplitude.} \vspace{-0mm}
	\label{fig:device}
\end{figure}

The DUT was built by sputtering 30 nm of T-shaped Pt and Au metal contacts ($\sim$ 1.5 mm width and 2.0 mm long) on top of the surface of the film, arranged in a planar structure. A schematic representation of the DUT is depicted in Fig.~\ref{fig:device} a).

%fig1

The two central bars of the Ts were located facing each other and are about 0.1 mm apart. This geometry was chosen to allow maximum exposure of the electrically active area of the DUT (where most of the voltage drops) preventing that any element block or affect the penetration of incident protons through it. With the same purpose in mind, Cu leads were carefully fixed over the pads by using silver paint, avoiding direct contact with the YBCO film and away from the active area. The Pt contact was labeled arbitrarily as ``+", while the one made with Au was used as ground (``-") pad. The choice of different metals as electrodes is based on previous results~\cite{Schulman13}.  As will be described later, this feature produce a memory device with essentially only one active interface. Finally, the DUT was placed inside a small outline integrated circuit (SOIC)-16 package for later integration.\\

Initially, the DUT was prepared in such a way to induce the resistive switching (reset) of the Au-YBCO (``-") interface from a low resistance state (LRS) to a high resistance state (HRS), so as to highlight its non-linear behavior. This condition makes it more sensitive to the microscopic changes that could occur due to irradiation than it would be if it behaves as an ohmic and more conducting interface. By using a B2902B Agilent source-measurement unit (SMU), we applied a series of 10 ms resetting positive current pulses (20 mA to 50 mA) at room temperature to increase the resistance of the Au-YBCO interface. The process is depicted in Fig.~\ref{fig:device} b). Note that the remanent resistance (R$_{rem}$) of the DUT, measured at two terminals with 0.5 mA, increased from 200 $\Omega$ to approximately 800 $\Omega$. The stability of the final state was then checked by measuring its IV characteristics, by applying 10 ms resetting current pulses of up to 15 mA as well as small reading pulses (delayed 1 s with respect to the writing pulses) to control the remanent value. As displayed in Fig.~\ref{fig:device} c), the measured instantaneous resistance (R$_{inst}$) and R$_{rem}$ show the non-linear behavior and the absence of RS in that current range, respectively. \\

%IV_01b_coninset_gamma.eps
\begin{figure} [h]
	\vspace{0mm}
	\centerline{\includegraphics[width=5in]{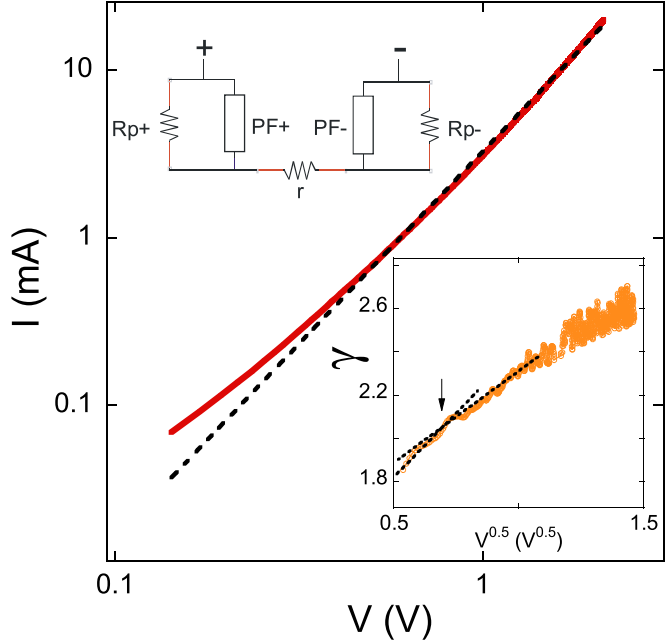}}
	%\centerline{\includegraphics[width=6.7cm,height=5.7cm]{R4wycontactos4.eps}}
	\vspace{-0mm}\caption{(Color online) IV characteristics of the YBCO-based device. The black dashed line is the best power law fit of the experimental curve (red line) showing that this law does not fit well the experimental data. As shown in the lower inset, the power exponent $\gamma$ is far from being constant (power law) and helps to reveal details of the IV dependence. The small change in the slope of the $\gamma$ dependence on V$^{0.5}$ (indicated with a black arrow) is showing the existence of two non-linear Poole-Frenkel regimes (dashed lines added as guides to the eye). The upper inset shows a possible circuit representation derived from previous publications and the $\gamma$ curve analysis. $r$ is the bulk YBCO resistor, PF$^{\pm}$ state for the non-linear Poole-Frenkel elements, Rp${\pm}$ represent the ohmic resistors in parallel, for the ``+" and ``-" contacts, respectively. } \vspace{-0mm}
	\label{fig:IV_LBT}
\end{figure}

Fig.~\ref{fig:IV_LBT} shows the IV characteristics of the DUT together with its $\gamma$ power exponent representation ($\gamma=dLn(I)/dLn(V)$ plotted as a function of $V^{1/2}$) and its proposed equivalent circuit model.  The $\gamma$ representation is a very helpful graphical method~\cite{Acha17} that serves to determine the existence of different transport mechanisms present in an interface, especially when there is more than one~\cite{Acha16,Acevedo17,Ghenzi19}. As can be observed in the $\gamma$ curve, two regions with linear behavior and having slightly different slopes (see the arrow and dashed lines) can be determined. Within the framework of the $\gamma$ representation, a linear behavior that extrapolates to 1 is a clear evidence of a Poole-Frenkel (PF) emission. The PF conduction is a bulk mechanism associated with the existence of traps in the YBCO oxide. Electric field and/or thermal fluctuations de-trap carriers, promoting them to the conduction bands and allowing the existence of a current in the dielectric material~\cite{Sze06,Chiu14}.  The appearance of two slopes is associated with the fact that each interface of our DUT is in a different resistance level. Throughout previous studies carried out on the same interfaces~\cite{Schulman12,Schulman13,Schulman15,Lanosa20}, a detailed circuit model has been established, which is depicted in the upper inset of Fig.~\ref{fig:IV_LBT}. The inhomogeneous oxygen distribution in the oxide near the interfacial region determines the existence of more than one conduction mechanism. In that sense, along with the PF emission, a parallel ohmic channel is developed. Finally, both interfacial regions are connected through the bulk resistance of the sample ($r \sim$ 1~$\Omega$), whose contribution to the total resistance of the DUT can be neglected. \\

To analyze the effects of the 10 MeV proton irradiation~\cite{SolarRadHandbook} on the electrical properties of the DUT, the IV characteristics were measured for the pristine state and after each proton irradiation interval. The irradiations were performed at room temperature in the heavy-ion tandem accelerator of the Tandar facility~\cite{Tandar} at the EDRA irradiation line (Test of Radiation and Environmental Damage)~\cite{Ibarra19}.  The DUT was electrically characterized {\textit{in situ}}, soldered onto one LabOSat-01 (LS-01) board~\cite{Barella16}. This board was specifically designed to electrically test electronic devices in hostile environments such as outer space, nuclear reactors, etc. In fact, LS-01 has been successfully used for the characterization of many TiO$_2$ and La$_{1/3}$Ca$_{2/3}$MnO$_3$ samples onboard low Earth orbit satellites~\cite{Sanca17,Barella19}. Besides, a similar in-orbit experiment was recently performed on YBCO-based ReRAM devices~\cite{Acha20}. \\

\begin{figure} [htb]
	\vspace{-2mm}
	\centerline{\includegraphics[width=6.0in]{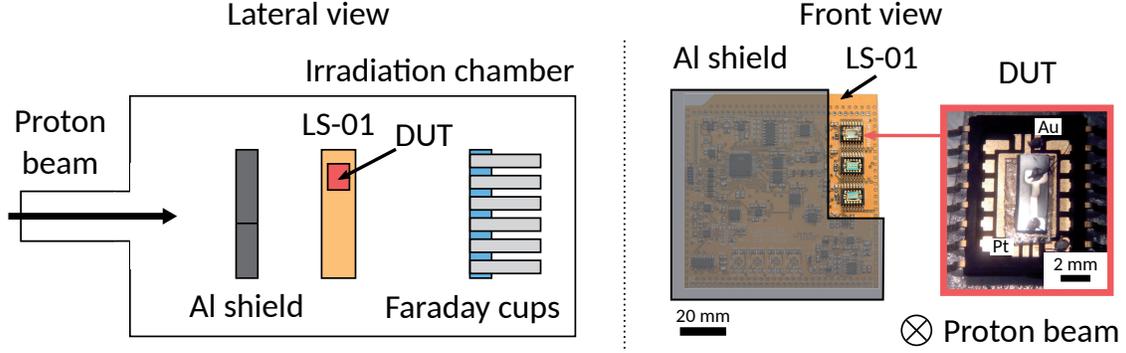}}
	%\centerline{\includegraphics[width=6.7cm,height=5.7cm]{R4wycontactos4.eps}}
	\vspace{-0mm}\caption{(Color online) Left) Schematic of the lateral view of the irradiation chamber showing the location of the Al shield, the LS-01 board, the detector and the direction of the incident proton beam. Right) Front view of the LS-01 board. The Al shield leaves exposed three SOIC-16 packages where samples (i.e., DUTs) may be contained (see Ref.~\cite{Barella16})]. The location of the YBCO DUT is pointed out with an arrow. The photograph shows the DUT placed inside the SOIC-16 package; notice its T-shaped metallic electrodes. } 
	\vspace{-0mm}
	\label{fig:chamber}
\end{figure}

To measure the IV characteristics of the DUT, LS-01 was configured to apply 100 ms current pulses while measuring the corresponding voltage drop,  performing a complete loop from 0$\rightarrow$-1.2 mA$\rightarrow$1.2 mA$\rightarrow$0 mA in steps of 15 $\mu$A. For clarity, considering that no rectifying behavior was observed and that similar results were obtained independently of the applied polarity, only data of the first positive quadrant are presented here. LS-01 with the DUT was placed inside the vacuum chamber of the irradiation line (see Fig.~\ref{fig:chamber}). As the proton beam waist is 8 cm long, an aluminum shield was used to protect the electronic circuits of LS-01 from radiation damage. The shield had a small aperture to allow protons to hit the DUT. To monitor the stability of the beam and to estimate the proton fluence, conventional Faraday cups located behind LS-01 were used to periodically monitor the beam flux. LS-01 was controlled remotely, far from the irradiation line, using a communication network.  The applied measurement and irradiation protocol is listed in Table~\ref{tab:dosis}.

%fig3

\begin{table}[h!]
	\centering
	\caption{Detail of the electrical measurements and their corresponding irradiation fluences. Temperature measured by LS-01. (Uncertainties of the reported values  are lower than 5\% for Flux or Fluence and $\pm$ 1 $^{\circ}$C  for Temperature.)}
	\vspace{0.5cm}
	\footnotesize{
		\begin{tabular}{||cccccc||}
		\hline \hline
			\textbf{Interva}l & \textbf{Duration} & \textbf{Flux} & \textbf{Fluence} & \textbf{Cumulative Fluence} & \textbf{Temperature} \\
			& (s) & ($\times$10$^{8}$ p/cm$^{2}$s) & (p/cm$^{2}$) & ($\times10^{9}$ p/cm$^{2}$) & ($^{\circ}$C)  \\ \hline \hline
		
			& \multicolumn{4}{c}{\textbf{LS-01 test}}		& 31		\\ \hline
			1 & 9 	& 3.12 & 2.8$\times$10$^{9}$ 	& 2.8 &		\\ \hline
			& \multicolumn{4}{c}{\textbf{LS-01 test}}  & 31				\\ \hline
			2 & 54  	& 3.30 & 17.8$\times$10$^{9}$ 	& 20.6	& \\ \hline
			& \multicolumn{4}{c}{\textbf{LS-01 test}}   & 32        			\\ \hline
			3 & 180 	& 3.34 & 6.0$\times$10$^{10}$ 	& 80.8    &  	\\ \hline
			& \multicolumn{4}{c}{\textbf{LS-01 test}}    & 32 				\\ \hline
			4 &  600  	& 3.57 & 2.1$\times$10$^{11}$ 	& 295  &    	\\ \hline
			& \multicolumn{4}{c}{\textbf{LS-01 test}}	& 33			\\ \hline
			& \multicolumn{4}{c}{30 min relaxation}  &	          			\\ \hline
			& \multicolumn{4}{c}{\textbf{LS-01 test}}	& 32		           			\\ 
		
			\hline \hline
		\end{tabular}
	}
	\label{tab:dosis}
\end{table}

\subsection{Monte Carlo TRIM simulations}
To gain insight into the effects produced by protons on the Au-YBCO stack, we performed  Kinetic Monte Carlo simulations by using TRIM, a set of tools included in the SRIM 2013 (Stopping and Range of Ions in Matter) package~\cite{Ziegler10,SRIM}. In order to compute the whole damage produced by protons on YBCO, a two step simulation process was considered:\\
First, an \textit{Ion Distribution and Quick Calculation of Damage} was performed using the Kinchin-Pease model analysis (run 1) with 1 million incident protons of 10 MeV. A stack of 30 nm gold film on top of a 1.5 $\mu$m YBCO film was used as target material. Notice that the YBCO film in the simulation is ten times thicker than in the actual sample. The reason for this is discussed below. Primary Knock-on Atoms (PKAs) spectrum~\cite{Marwick75,Gilbert15} was calculated for the whole stack. Energy loss by protons is negligible across the thicker YBCO layer. Only 0.02 \% of their initial energy is lost during transport. Having used an YBCO layer ten times as thick as the actual one allowed us to collect ten times more data without affecting the underlying physics of the process. The result is equivalent to perform a simulation with 10 million protons, a quantity that is unpractical to use with TRIM code. \\
Second, in order to assess additional damage produced by the high density cascades of displaced ions, the obtained PKAs spectrum was used to perform a second simulation (run 2) to estimate the damage of the full cascade.

\section{RESULTS AND DISCUSSION}

\begin{figure} [htb]
	\vspace{0mm}
	\centerline{\includegraphics[width=6.5in]{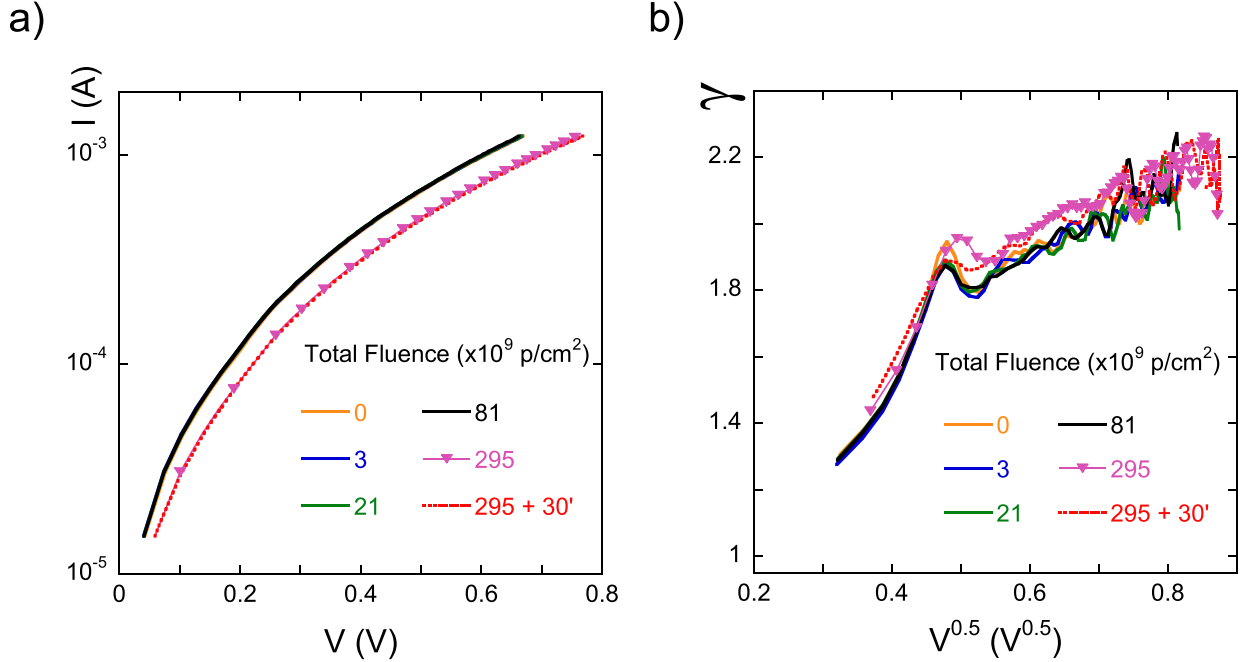}}
	%\centerline{\includegraphics[width=6.7cm,height=5.7cm]{R4wycontactos4.eps}}
	\vspace{-1mm}\caption{(Color online) a) IV characteristics of the YBCO-based device for different proton cumulated fluences. Curves corresponding to up to 81$\cdot$10$^9$ p/cm$^2$ fall one on top of the other. Only the 295$\cdot$10$^9$ p/cm$^2$ curve displays a clear shift to a less conducting device is observed. A measurement taken 30 min after the last irradiation interval (pink dashed curve) shows no evidence of relaxation. b) $\gamma$ representation of the same measurements presented in a).} \vspace{-0mm}
	\label{fig:IVexp}
\end{figure}

Fig.~\ref{fig:IVexp} a) and b) show the IV characteristics measured by the LS-01 board and its $\gamma$ curves for the total applied proton fluences, respectively. No changes in the IV characteristics were noticed with fluences of up to $81 \cdot 10^{9}$ p/cm$^{2}$. All curves fall on top of each other, making it impossible to distinguish them. The same occurs for the $\gamma$ curves. Thus, as no microscopic changes were noticed in the electrical characteristics, no changes in the memory operating parameters (e.g., V$_{set}$, V$_{reset}$ or the resistance change in the RHSL) nor in its performance should be expected.

%IV_y_gamma_dosexx.eps

This is a promising result, as it shows how tolerant the devices are to one of the most abundant ionizing sources in LEO. A straight comparison with operational conditions in outer space is a complex task. Proton fluence depends on orbital parameters and solar activity as well as on the spacecraft or satellite shielding. For instance, according to SRIM simulations, 10 MeV protons are likely to be stopped by a 2 mm thick aluminum shield. Although protons with energy less than 20 MeV will not reach the interior of the spacecraft, high energy protons would pass through the shield inducing radiation damage into the spacecraft electronics. Nevertheless, to make a comparison with our results, simulations using the Space Environment Information System tool~\cite{Spenvis04} developed by the European Space Agency were carried out. In terms of particle fluence, for a 1 yr mission at a 500 km altitude orbit, the total fluence of 10 MeV protons during 2016 solar maximum is in the order of 10$^{9}$ p/cm$^{2}$. Thus, our results indicate that YBCO-based memory devices could stand 80 times the proton fluence found in LEO. \\

A very clear shift of the IV curve to a lower conducting state occurs only for the highest applied fluence (295$\cdot$10$^9$ p/cm$^2$). During the next 30 minutes, no relaxation of the produced effects was observed. Measurements made by the LS-01 board show more clearly the existence of two slopes and a cross-over voltage in the $\gamma$ curves than those obtained by the Agilent SMU. This is due to the fact that the voltage sweep that was performed with the LS-01 board starting from lower values showed more clearly the low-voltage ohmic contribution ($\gamma \sim 1$) due to the existence of resistor in parallel with the PF element. Here again, the 295$\cdot$10$^9$ p/cm$^2$ fluence shifts the curves to a higher non-linear behavior, which were tested to remain stable for the next 30 minutes. \\

A better understanding of the microscopic effects produced by the proton irradiation can be obtained by analyzing quantitatively the changes induced in the electrical transport properties. By considering the equivalent circuit model, the IV characteristics of each interface ($``+"$ and $``-"$) can be described in terms of the PF ($I_{PF}^{\pm}$) and its parallel ohmic leak ($I_{R_p}^{\pm}$) currents, as stated  by the following equations:

\begin{equation}
\label{eq:I1} 
I =  I_{PF}^{\pm} + I_{R_p}^{\pm}~, 
\end{equation}

\begin{equation}
\label{eq:V} 
V = V_{PF}^{+} +  V_{PF}^{-} ~,
\end{equation}

\begin{equation}
\label{eq:PF} 
I_{PF}^{\pm} = A^{\pm}~V_{PF}^{\pm} ~ \exp[B^{\pm}
(V_{PF}^{\pm})^{1/2}]  \hspace{2mm} {\normalfont{and}} \hspace{2mm} I_{R_p}^{\pm} = \frac{V_{PF}^{\pm}}{R_p^{\pm}}~,
\end{equation}

\noindent with
\begin{equation}
\label{eq:AyB}
A^{\pm}=\frac{\exp[-E_{Trap}^{\pm}/(k_B T)]}{R_0^{\pm}} , \hspace{0.5mm}  B^{\pm} = \frac{q^{3/2}}{k_B T (\pi \epsilon'^{\pm} d^{\pm})^{1/2}}~,
\end{equation}

\noindent where $R_0^{\pm}$ is a pre-factor associated with the geometry of the conducting path, the
electronic drift mobility ($\mu$) and the density of states in the
conduction band. $E_{Trap}^{\pm}$ is the trap energy level, $k_B$ the Boltzmann
constant, $q$ the electron charge, $\epsilon'^{\pm}$ the real part of the dielectric constant of the oxide and $d^{\pm}$ the interfacial thickness where most of the voltage drops (for each interface ${\pm}$). 

\begin{figure} [htb]
	\vspace{0mm}
	\centerline{\includegraphics[width=6.5in]{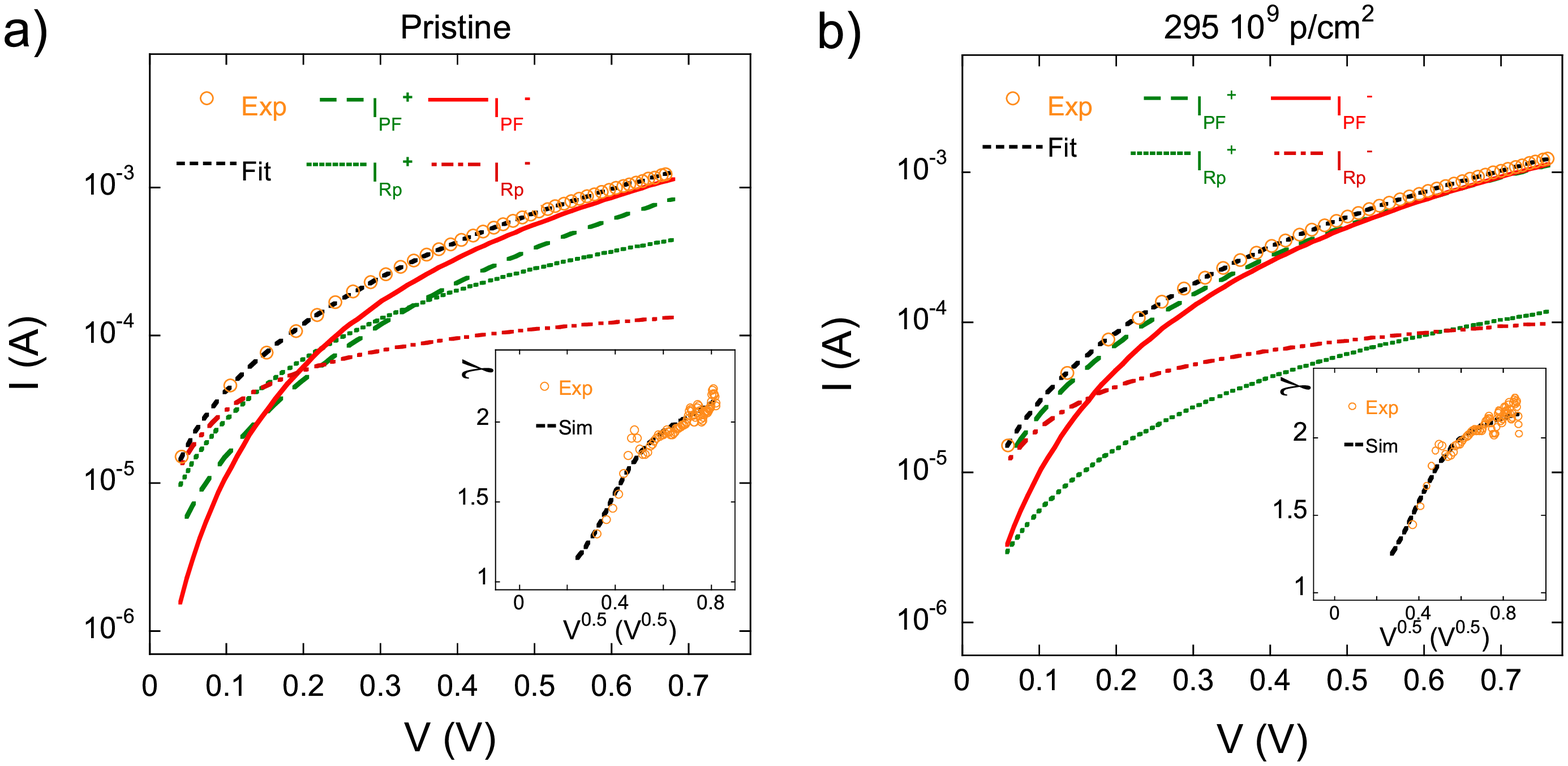}}
	%\centerline{\includegraphics[width=6.7cm,height=5.7cm]{R4wycontactos4.eps}}
	\vspace{-2mm}\caption{(Color online) Fitting curves for the IV characteristics showing the contribution of each interfacial element for the a) pristine and b) proton irradiated (295$\cdot$10$^{9}$ p/cm$^2$) YBCO-based device. Both insets show the experimental and the fitted $\gamma$ vs V$^{1/2}$ dependence.} 
	\vspace{-0mm}
	\label{fig:IVexpysim}
\end{figure}

By replacing the expressions of Eq.~2 and Eq.~3 in Eq.~1 and by equating the total current of each interface, the implicit and non-linear Eq.~1 was solved numerically by applying a generalized reduced gradient method. The purpose of this was to obtain the parameters $A^{\pm}$, $B^{\pm}$ and $R_p^{\pm}$ for the pristine and the most irradiated DUT. Different starting values for these parameters were tested allowing us to obtain the solution with the lower residuals. We have checked that the obtained parameters for this solution (within a 10 \% interval) also generate the lower residuals when fitting the data of the negative quadrant (not presented here).
Our best results comparing the experimental and the calculated IV characteristics and their corresponding $\gamma$ representation are shown in Fig.~\ref{fig:IVexpysim}. The contribution to the current of each circuit element is also presented. The obtained parameters are summarized in Table~\ref{tab:ajustes}.

%IV_y_gamma_ajustes_compara.eps

\begin{table}[h]
	\caption{Fitting parameters derived from numerically solving
		Eq.~1 for the pristine and the proton irradiated (295$\cdot$10$^{9}$ p/cm$^2$) DUT. The ratio of the irradiated over the pristine values is also presented.}
	\vspace{0.4cm}
	\label{tab:ajustes}
	\begin{tabular}{||c||ccc|cc|cc|cc|cc|cc||}
		\hline  \hline
		\textbf{State}      && \large{$A^{+}$} && \large{$B^{+}$} && \large{$R_p^+$}  && \large{$A^-$} && \large{$B^{-}$} && \large{$R_p^-$}  & \vspace{-2mm}\\
		&& \textbf{($\mu$S)} && \textbf{($V^{-1/2}$)} && \textbf{($\Omega$)}  && \textbf{($\mu$S)} && \textbf{($V^{-1/2}$)} && \textbf{($\Omega$)} & \\ \hline \hline
		\textbf{Pristine} & &      490     & &    2.7    & &    790      & &  6.9 && 10.8 && 2500  &        \\
		\textbf{Irradiated} & &     1200    & &    1.7      & &     2900       & &    9.5  &&  8.7 && 4300 &  \\
		\textbf{Ratio} & &     2.4    & &    0.6      & &     3.7       & &    1.4  &&  0.8 && 1.7 &       \\ \hline \hline
	\end{tabular}
\end{table}

The best fits of the IV characteristics excellently reproduce the details of the experimental data. It can be observed that for the pristine sample, leak resistors make an important contribution to the total current at low and intermediate voltages, while proton irradiation decreases this contribution. As in a series configuration the overall non-linear behavior is determined by the less conducting PF element, the observed change in the slope of the $\gamma$ curve ($\sim B$ for a single PF emission) is related to the poorest conduction of the PF$^-$ element at low voltages (with $B^- > B^+$). In contrast, the roles are interchanged at higher voltages, determining a $\gamma$ limited by the PF$^+$ element, with a lower slope for this voltage region. This effect is less pronounced for the irradiated DUT.  The peak observed in the experimental $\gamma$ was not reproduced by our simulations, as we obtain a smooth cross-over from one regime to another. The experimental peak can be related to a more complex equivalent circuit that may include a small contribution of a Fowler-Nordheim element~\cite{Sze06}, not considered for simplicity.
It can be noticed that proton radiation increases the resistance of the ohmic resistors while it favours the PF emission. These effects are more evident at the Pt-YBCO $``+"$ (i.e., the most conducting and oxygen-rich) interface than at the Au-YBCO $``-"$ one. The same occurs with the PF emission. \\

\begin{figure} [t]
	\vspace{0mm}
	\centerline{\includegraphics[width=6.5in]{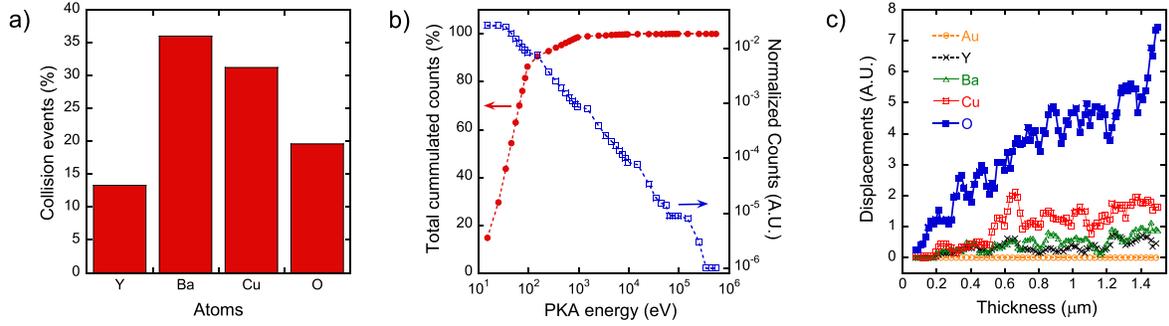}}
	%\centerline{\includegraphics[width=6.7cm,height=5.7cm]{R4wycontactos4.eps}}
	\vspace{-5mm}\caption{(Color online) a) PKAs for each element of the YBCO film and b) PKA spectrum normalized for the total number of incident protons, both obtained for run 1. c) Number of displacements  by element as a function of the thickness of the film, obtained after run 2 (displacements produced by cascades are included) .} 
	\vspace{-0mm}
	\label{fig:TRIM}
\end{figure}

The results obtained for Monte Carlo TRIM simulations are presented in Fig.~\ref{fig:TRIM}. The PKAs resulting from run 1 computes a total of $\sim$26,000 events. As it is shown in Figure~\ref{fig:TRIM}a, most displaced atoms are Ba and Cu, while almost 20 \% of the displacements involve O displacements and only 14 \% are associated with Y. It should be noted here that since the energy (E) of the incident protons does not vary throughout the thickness of the film, then the Rutherford cross section (proportional to E$^{-2}$) should be constant along all the paths and the probability of generating defects would be uniform throughout the film. The PKAs spectrum is shown in Fig.~\ref{fig:TRIM}b. As can be observed, almost 95 \% of  the PKAs have energies below 100 eV. Thus most of the displaced ions produced during the cascade will not decay as rearrangements of the lattice, so they survive as permanent defects.

%TRIM.eps

The results of the Full Cascade of run 2, presented in Fig.~\ref{fig:TRIM} c), show the overall displacements derived from the cascades involving all the ions. It can be observed that, independently of the thickness of the film, the majority of the displacements now correspond to oxygen ions. Although this should be taken as a qualitative result due to the ideal conditions at which TRIM simulations are performed, the observed trend seems to be relevant for the resistive switching mechanism, as oxygen content dominates the resistive state. \\

In this way, both our IV measurements and TRIM simulations indicate that protons mainly produce the knock-on of oxygens, displacing them from their structural site. As this generation of point defects would be more likely in the near optimally oxygenated YBCO, where oxygen is principally located, the areas presenting the ohmic conduction would be more affected, in accordance to the changes produced by proton irradiation on the fitting parameters (see the ratios presented in Table~\ref{tab:ajustes}). 

\begin{figure} [t]
	\vspace{-10mm}
	\centerline{\includegraphics[width=6.5in]{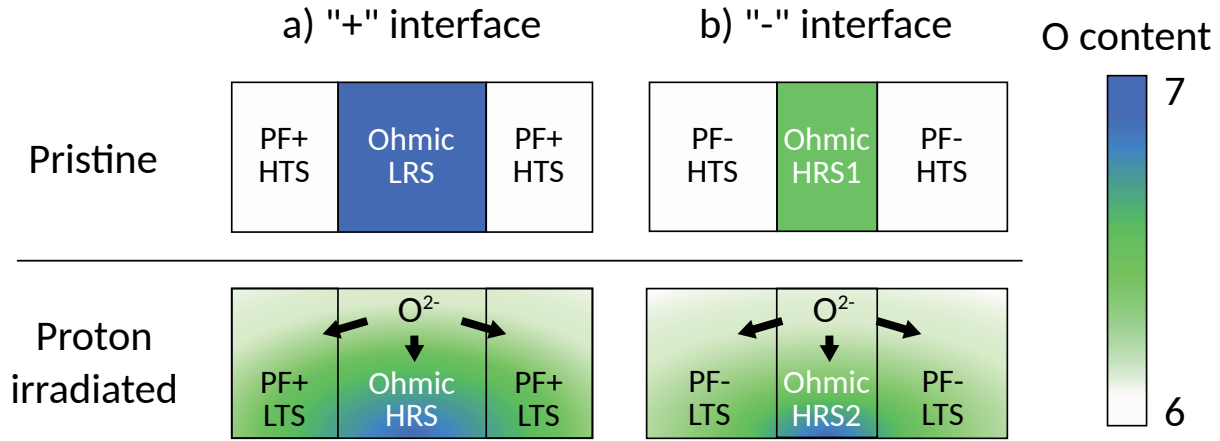}}
	%\centerline{\includegraphics[width=6.7cm,height=5.7cm]{R4wycontactos4.eps}}
	\vspace{-0mm}\caption{(Color online) Schematic representation of the coexistence of ohmic and PF regions in the oxide, close to the surface for the a)  $``+"$ Pt-YBCO and b) $``-"$ Au-YBCO interfaces, for the pristine and the proton-irradiated device. The evolution of the electrical transport parameters suggest that oxygen is knocked on from its position, inducing a migration from areas with high oxygen content to those with lower oxygen content, where it becomes trapped. (LTS and HTS, low trap and high trap state, respectively).} 
	\vspace{-0mm}
	\label{fig:schematic}
\end{figure}

This situation is schematically represented in Fig.~\ref{fig:schematic}. It can be observed that, when displaced, oxygens will diffuse until they are trapped, with a high probability of this happening precisely in the traps that are located in the surrounding PF areas. This is what can be inferred from the observed increase of the parameter A, which can be an indicator of a decrease in the energy of the traps (see Eq.~4). This is also consistent with the fact that the traps are related to oxygen vacancies that may get filled by the displaced oxygens. If the number of trapped oxygens is large, some conductive nano islands must be formed in the former PF zones, which is also reflected in the reduction of parameter B. Indeed, the mixture of metallic and insulating zones will generate the formation of nanocapacitors, which due to the Maxwell-Wagner effect~\cite{VonHippel62,Lunkenheimer10}, will increase the dielectric constant and therefore reduce the B parameter (see Eq.~4).  

%schematic_oxygen_knockout.eps

Thus, our studies confirm which is the specific damage caused on YBCO by its irradiation with 10 MeV protons. This question was addressed in previous studies that tried to unveil which were the main defects generated by protons of this energy range on YBCO, but were not conclusive. A previous report~\cite{Kirk93}, based on transmission electron microscopy and magnetization measurements at low temperature, states that protons mainly generate non-visible point defects, but the nature of these defects (i.e., whether they are related to oxygen or copper displacements) was not determined. Other studies, by performing molecular dynamic simulations of radiation-induced ion displacements in YBCO, indicated that the preferential radiation damage occurs in the oxygen sublattice of YBCO~\cite{Cui92}. More recently, studies combining micro-Raman studies with proton-damage simulations~\cite{Li19} arrive at the same conclusions: low energy protons impacting on YBCO mainly generate O displacements. These results give a clue regarding the strategy to follow in order to achieve memories hardened to proton irrradiation. Indeed, it would be expected that interfaces based on a YBCO with poor oxygen content would be less damaged by protons, thus allowing to extend the maximum fluence for which no changes in their electrical properties are detected.

\section{CONCLUSIONS}
We have characterized the effects of proton irradiation (10 MeV) on the IV characteristics of
a metal-YBCO thin film test structure at room temperature. Up to fluences as high as 81$\cdot$10$^9$ p/cm$^2$ no effects were observed, giving a clear indication that ReRAM-type memories based on perovskite oxides seem to be suitable for use under hostile environments with high proton radiation. An equivalent circuit model was proposed, which reproduces very well the details of the experimental IV characteristics and indicates the degree of inhomogeneity of the oxygen distribution within YBCO, near the interfaces. We also performed a deep analysis of the microscopic effects produced for a higher fluence of 295$\cdot$10$^9$ p/cm$^2$, which, in accordance with Monte Carlo TRIM simulations, determine an scenario where protons increase the overall resistance of the DUT, mainly by inducing oxygen migration from oxygen-rich areas (ohmic conduction) to oxygen-poor areas (PF emission), where it ends up being trapped, locally modifying the electrical transport. Although our YBCO-based test structure is far from being a real memory device, our studies may serve as an initial proof of concept for validating the use of perovskite oxides to build ReRAM devices for a particular use. Indeed, our results indicate that, in the quest for materials intended to optimize the operation of electronic memories in harsh environments,  low-oxygen-content YBCO could occupy a prominent place in the design of memories whose main requirement is to remain operational under high irradiation fluence of protons.

\section{ACKNOWLEDGEMENTS}
We would like to acknowledge financial support by CONICET grants PIP 11220150100322 \& 11220150100653CO, ANPCyT grants PICT 2016-0966 \& 2017-0984 and UBACyT 20020170100284BA
(2018-2020). Jenny and Antti Wihuri Foundation is also acknowledged
for financial support. We thank Eng. E. P\'erez Wodtke for
its technical collaboration as well as the Tandar operators team of CNEA for their assistance during irradiations.

%\pagebreak
%\newpage
%\bibliography{bibRRAM}

\end{document}